\documentclass[a4paper,11pt]{article}
\usepackage{jcappub}
\usepackage{amsmath,amssymb,amsthm}

\usepackage{mathrsfs}     
\usepackage{empheq}       
\usepackage{mathtools}    
\usepackage{bm}           
\usepackage{dcolumn}      
\usepackage{cleveref}     
\usepackage{soul}         
\usepackage{comment}      
\usepackage{booktabs}
\usepackage{array}         
\usepackage[table]{xcolor} 
\usepackage{colortbl}      
\definecolor{tabcol0}{RGB}{242,242,242}
\definecolor{tabcol}{RGB}{225,246,250}
\usepackage{fontawesome5} 
\usepackage{orcidlink}    

\definecolor{darkblue}{rgb}{0.0,0.0,0.5}
\hypersetup{
    linkcolor=darkblue,
    citecolor=blue,
    urlcolor=darkblue,
}

\newcommand{\be}{\begin{equation}}
\newcommand{\ee}{\end{equation}}
\newcommand{\bs}{\begin{split}}
\newcommand{\bea}{\begin{eqnarray}}
\newcommand{\eea}{\end{eqnarray}}
\newcommand{\al}{\alpha}
\newcommand{\alb}{\alpha_B}
\newcommand{\alk}{\alpha_K}

\newcommand{\Om}{\Omega_m}
\newcommand{\om}{\Omega_m}
\newcommand{\ode}{\Omega_{\rm de}} 
\newcommand{\odei}{\Omega_{{\rm de},i}}
\newcommand{\odet}{\Omega_{{\rm de},0}}
\newcommand{\wde}{w_{\rm de}}
\newcommand{\wtot}{w_{\rm tot}}

\newcommand{\rde}{\rho_{\rm de}}
\newcommand{\pde}{P_{\rm de}}

\newcommand{\lcdm}{$\Lambda$CDM} 
\newcommand{\wowa}{$w_0w_a$CDM}
\newcommand{\gm}{G_{\rm matter}}
\newcommand{\gl}{G_{\rm light}}
\newcommand{\geff}{G_{\rm eff}}

\newcommand{\mpl}{M_{\rm Pl}^2}

\newcommand{\lam}{\lambda}

\newcommand{\wl}{w_{\rm fut}}

\newcommand{\we}{w_{\rm early}}

\newcommand{\github}[1]{\href{#1}{\faGithub}}

\title{Viable Phantom Crossing in Softly Broken Shift Symmetric Gravity} 

\author[a]{Rodrigo Calder\'on\orcidlink{0000-0002-8215-7292}}
\author[b]{Eric V.\ Linder\orcidlink{0000-0001-5536-9241}}

\affiliation[a]{CEICO, Institute of Physics of the Czech Academy of Sciences,
  Prague, Czech Republic}
\affiliation[b]{Berkeley Center for Cosmological Physics \& Berkeley Lab,
  University of California, Berkeley, CA 94720, USA}

\emailAdd{calderon@fzu.cz}
\emailAdd{evlinder@lbl.gov}

\abstract{We demonstrate a viable fit to current cosmological 
data from a simple Horndeski modified gravity theory that 
softly breaks shift symmetry. This SBSS model crosses the phantom 
divide $w=-1$ at $z\approx0.5$ and fits observational data (baryon acoustic oscillation and supernova distances, and cosmic microwave background power spectra) 
better than $w_0w_a$CDM, by 
$\Delta\chi^2\approx-3$ for the same number of effective parameters, while providing a physical foundation for cosmic acceleration. 
The theory is healthy, has no cosmological constant or 
potential, and 
has substantial 
protection against high energy quantum loop corrections.}

\begin{document}

\maketitle

\section{Introduction}

Cosmological data currently gives a consistent picture \cite{desi2,desi2de} 
where the deceleration 
parameter $q(z)$ is less positive than in \lcdm\ at redshift $z\approx1$ and less 
negative at redshifts $z\lesssim0.5$. This corresponds to containing a component 
more strongly accelerating than a cosmological constant $\Lambda$ 
(possessing effective dark energy equation of state ratio $w(z)<-1$) 
at $z\approx1$ then more weakly accelerating ($w(z)>-1$) at $z\lesssim0.5$. 

Within general relativity, a minimally coupled scalar field cannot achieve 
such a crossing of $w=-1$, referred to as the phantom divide. This motivates 
either looking outside general relativity, or abandoning minimal coupling. Here 
we focus on modified gravity, specifically Horndeski gravity, while keeping 
minimal coupling of all components. This gives a modified gravity consisting 
of the usual Einstein-Hilbert action, including a scalar field sector (involving 
a kinetic/potential term, called $K$ or sometimes $G_2$), plus a braiding term 
(called $G_3$). 

There is little guidance for how these terms should depend on the field and its 
derivative, so we adopt the following guideposts (these are not formally derived 
requirements but rather desiderata): 
\begin{itemize} 
\item Simplicity -- as few terms, and effective parameters, as possible; 
\item Protection -- adhering to shift symmetry, or other protections against 
quantum radiative corrections, to a large extent and in particular at high 
energies in the early universe;  
\item Naturality -- no cosmological constant (whose magnitude is problematic), 
or any potential; the modified gravity effects of 
braiding and noncanonical kinetic structure alone dominate the key aspects of 
evolution (``naturality'' indicates that this is not intended as a formal 
naturalness claim). 
\end{itemize} 

Section~\ref{sec:model} presents the modified gravity model 
respecting these guideposts that we explore. Section~\ref{sec:eom} presents the key 
cosmological quantities and the numerical evolution equations corresponding 
to them. 
We derive analytically the dynamical fixed points, e.g.\ the 
dark energy fate, in Section~\ref{sec:wfinal}. 
Numerical solutions are evaluated and compared to observational data in 
Section~\ref{sec:results}. We interpret the results and discuss conclusions in Section~\ref{sec:concl}.

\section{A Shift Symmetry Motivated Model} \label{sec:model} 

The general Horndeski Lagrangian can be simplified by two assumptions, 
which we adopt here. One is the luminality assumption -- that the progagation 
speed of gravitational waves is the speed of light. This is motivated by 
observations of GW170817 \cite{170817} but can be weakened by consideration that 
the measurements are near the theory's strong coupling scale \cite{melville} 
and at a single redshift, $z\approx0$. Nevertheless, until further measurements 
occur, from gravitational wave events with electromagnetic counterparts, or at very 
different frequencies, e.g.\ from pulsar timing arrays or spaceborne 
interferometers, we will treat the luminality in the most straightforward 
manner. This implies lack of a $G_5$ Lagrangian term and that any $G_4$ term 
does not depend on field derivatives. The second assumption is minimal gravitational 
coupling, i.e.\ the Ricci scalar coupling to matter or a scalar field is not 
modified. 

This brings the Lagrangian to the form \cite{1008.0048,Pujolas:2011he,Kobayashi:2010cm}
\be 
{\mathcal L}=\frac{1}{2}\,R+K(\phi,X)-G_3(\phi,X)\,\Box\phi\ . 
\ee 
That is, $G_4(\phi,X)=1/2$, where we have normalized the Planck mass to 
$\mpl=1$. Here $X=-(1/2)\phi^{,\mu}\phi_{,\mu}=(1/2)\dot\phi^2$. 
In a purely shift symmetric theory, neither $K$ nor $G_3$ would have 
dependence on the field value $\phi$. Such a theory was 
previously thought unable to cross 
the phantom divide from below ($w<-1\to w>-1$) \cite{tsuji}, but recent work demonstrated 
that it can under special circumstances \cite{Calderon:2026vgr} -- 
but these do not yield a cosmology in good agreement with 
observations (e.g.\ $z\lesssim1$ phantom crossing). 

Instead we will break shift symmetry by adding field 
dependent terms, but in a particular way. Following our 
guideposts we write 
\be 
{\mathcal L}=\frac{1}{2}\,R+V_k(\phi)\,\kappa(X)-V_g(\phi)\,\gamma(X)\,\Box\phi\ . \label{eq:actionkgsep}
\ee 
That is, we have treated the functions $K$ and $G_3$ as separable 
in field value and field derivatives. This is quite common in many 
modified gravity models (e.g.\ some generalized Galileons) but we do not claim any 
formal fundamental basis for this; we follow it due to the simplicity 
guidepost (and the protection guidepost as we will soon see). 
Simplicity motivates us to take a single term for the kinetic 
function and a single term for the braiding function, rather 
than multiple terms or a general infinite series. Protection 
motivates us to have the field dependent terms $V_k$ and $V_g$ 
very nearly constant at high redshift ($z\gg1$) so that shift 
symmetry nearly holds in the early universe. Finally, naturality motivates us to take 
the kinetic function separable multiplicatively rather than 
additively (i.e.\ $\kappa(X)+V_k(\phi)$);  again we do not want 
to include a cosmological constant or have a potential 
drive the cosmic evolution. (For broken shift symmetric models with an explicit  potential $V(\phi)$, see e.g.\ \cite{tsuji,2509.17586}.) 
We freely acknowledge these are motivations, not requirements; 
certainly with infinite series expansions for the action terms one could likely 
obtain nearly any evolutionary behavior, but the data will not 
be able to map out more than very simple functional forms. 
We want to see whether we can succeed in fitting cosmological data while adhering as closely as possible to our guideposts. 

Thus we have two functions of $\phi$ and two functions of $X$ 
to deal with. Could we simplify this further? If we carry out a 
field redefinition to remove the $\phi$ dependence from the 
kinetic term, the redefined $\Box\phi$ braiding term will 
generate an additional kinetic term with both field value and 
field derivative dependence -- and in precisely the multiplicative 
separable manner we have assumed (another reason for this 
assumption). If we carry out a 
field redefinition to remove the $\phi$ dependence from the 
braiding term, again an additional kinetic term 
$K^{\rm new}(\phi,X)$ is generated (compare \cite{Hallam:2026qsk}). Thus we cannot get simpler 
than two functions of $\phi$ and two functions of $X$ unless we 
restrict the theory by hand, e.g.\ fixing the functions. 
See Appendix~\ref{sec:redef} for further details on field 
redefinition effects.

\section{Cosmological Quantities and Evolution Equations} \label{sec:eom}

\subsection{Modified Gravity Model and Effective Dark Energy} 
\label{sec:coseqs} 

We can define the effective dark energy density and pressure as 
\bea 
\rde&=& 2XK_X-K+6H\dot\phi g-2XG_{3\phi}\\ 
\pde&=& K-2g\ddot\phi-2XG_{3\phi}\ , 
\eea 
where a subscript of $\phi$ or $X$ denotes a derivative with 
respect to that quantity, and $g\equiv XG_{3X}$. 
This leads to a dark energy equation of state 
\be 
w=\frac{K-2g\ddot\phi-2XG_{3\phi}}{2XK_X-K+6H\dot\phi g-2XG_{3\phi}}\ , 
\ee 
and a total equation of state 
\be 
\wtot=w\ode+\Omega_r/3\ , \label{eq:wtot}
\ee 
including radiation energy density $\Omega_r$, or equivalently a deceleration parameter 
\be 
q=\frac{1}{2}+\frac{3}{2}\wtot\ . 
\ee 

The Friedmann equations and field equation of motion can be written as  
\bea 
3H^2&=&\rho_m+\rde\ ,\\
-2\dot H&=&\rho_m+P_m+\rde+\pde\ ,\\
0&=&\ddot\phi\,\left[K_X+2XK_{XX}+6H\dot\phi g_X-2G_{3\phi}-2XG_{3\phi X}\right] \label{eq:fd}\\
&\qquad&\quad+3H\dot\phi \left(K_X-2G_{3\phi}+2XG_{3\phi X}\right)-K_\phi+2XK_{\phi X}-2XG_{3\phi\phi}+6g\left(\dot H+3H^2\right)\ ,\notag  
\eea 
with $\rho_m$ and $P_m$ the energy density and pressure of all components outside the dark energy sector. 
The field equation can be rewritten as a first order charge density 
evolution equation, with 
\cite{bellsaw} 
\bea 
C &\equiv& \dot\phi K_X+6Hg-2\dot\phi G_{3\phi}\\ 
\dot C+3HC&=& K_\phi-2XG_{3\phi\phi}-2\ddot\phi XG_{3\phi X}\equiv {\mathcal P}_\phi\ . 
\eea 
When ${\mathcal P}_\phi$ is small compared to the other terms in 
the evolution equation then the scalar field charge density $C$ 
simply dilutes with the expansion of the universe, $C\sim a^{-3}$, 
conserving the 
charge itself. This is a characteristic of shift symmetric theories. 
Our theory is not shift symmetric, due to the field dependent terms, 
but by choosing $V_k(\phi)$ and $V_g(\phi)$ to be nearly constant 
in the early universe, then the ${\mathcal P}_\phi$ source term 
can be neglected there and the theory closely follows the shift symmetric 
characteristics at early times, thus 
providing quantum protection at high energies. 
We call this softly broken shift symmetry. 

To implement the softly broken shift symmetry, keeping ${\mathcal P}_\phi$ negligible and preserving 
charge conservation at early times, we adopt exponential forms for $V_k(\phi)$ and $V_g(\phi)$  
(inspired by the Asymptotic Cubic Galileon approach of \cite{Naidoo:2026umv}). 
This form also has the beneficial  
property of possessing some quantum protection in itself. 

For the $X$ dependence we take    
$K\sim X^n$ and $G_3\sim X^m$, with $n$ and $m$ constants. 
Taking monomials (besides the similarities to Galileons) is 
motivated by the fact that at early times 
{\it some\/} asymptotic dependence will dominate each one, and 
then for simplicity we assume the same functional dependence 
holds for all times. Otherwise the number of 
free parameters further increases, and without guidance to a particular functional form 
-- again our philosophy is to lead with simplicity 
until forced to complexify. That is, we explore whether this 
simplest model can successfully cross the phantom divide -- happily it can! 

Our modified gravity model, called Softly Broken Shift 
Symmetry gravity (SBSS), is therefore 
\bea 
K(\phi,X)&=&c_k\, e^{-\lam_k\phi}\,X^n\\ 
G_3(\phi,X)&=&c_g\, e^{-\lam_g\phi}\,X^m\ . 
\eea 
Note that $m=0$ would remove $X$ dependence from $G_3$, turning 
the braiding action term into a total derivative plus an added 
kinetic term $K^{\rm add}=-2G_{3\phi}X$; such a case is simply 
k-essence and we know it cannot cross the phantom divide, so we 
do not consider $m=0$. 
The constants $c_k$ and $c_g$ will be absorbed into the initial condition for the dark energy density $\odei$ and the initial fractional contribution $f$ of the braiding term (see \cite{paper1,Calderon:2026vgr}). 
Below we define $u$ as a logarithmic derivative of 
$e^{-\lam_g\phi}$ and $r\equiv\lam_k/\lam_g$.

\subsection{Dynamical Equations} \label{sec:eqsdyn} 

For numerical solution of the evolution we create an autonomous system of coupled 
first order differential equations, defining the dimensionless variables: 
\be  
x_k = \frac{K}{3H^2\mpl}\ , \qquad
x_g = \frac{6H\dot\phi\,g}{3H^2\mpl}\ , \qquad  
u = \frac{d\ln V_g}{d\ln a}=\frac{d\ln V_g}{d\ln\phi}\,\frac{d\ln\phi}{d\ln a} \ , 
\ee 

where we include $\mpl$ explicitly here to show the variables are 
dimensionless. Note that $d\ln V_k/d\ln a=ru$, with $r\equiv \lambda_k/\lambda_g$ a 
constant. We define 
the helper quantities 
\bea 
x_\phi &=& \frac{\ddot\phi}{3H\dot\phi} = \frac{X'}{6X}\ , 
\qquad x_H = \frac{d\ln H^2}{d\ln a} = -3(1+\wtot) \ , 
\eea 
where a prime denotes $d/d\ln a$. 

The full ODE system closes in $(x_k,x_g,u,\Omega_r)$:
\begin{equation}
\begin{aligned}\label{eq:ode_autonomous}
x_k'&=x_k\left[6n\,x_\phi-x_H+r\,u\right], &\qquad
x_g' &= x_g\left[\,3x_\phi(1+2m) - \frac{1}{2} x_H + u\,\right]\ ,\\
u'&=u\left[3x_\phi-\tfrac12x_H\right], &
\Omega_r'&=-\Omega_r\left[1-3w_{\rm tot}\right],
\end{aligned}
\end{equation} 
where $\wtot$ is given by Eq.~(\ref{eq:wtot}) and $x_\phi$ is algebraically determined from Eq.~(\ref{eq:alxphi}) below. 
The dark energy density and pressure can be written 
\begin{align}
\ode &= \frac{\rde}{3H^2\mpl} = (2n-1)\,x_k + x_g\!\left(1 - \frac{u}{6m}\right), \label{eq:odeu}\\
w\ode &= \frac{\pde}{3H^2\mpl} = x_k - x_g\!\left(x_\phi + \frac{u}{6m}\right)\ . \label{eq:wodeu} 
\end{align} 

We can write the braiding and kineticity property functions 
\bea 
\alpha_B &=& x_g \\ 
\alpha_K &=& 6n(2n-1)\,x_k + x_g\!\left(6m - \frac{1+m}{m}\,u\right)\ . 
\eea 
We use these to check the no ghost 
condition 
\be
\alpha\equiv\alk+\frac{3}{2}\,\alb^2\ge 0\ ,
\ee 
and the stability condition of having nonnegative sound speed squared of the scalar
perturbations,
\bea  
\alpha c_s^2&=& \left(1-\frac{\alb}{2}\right)\left(\alb-x_H\right)+\alb'-\frac{\rho_m+P_m}{H^2}\\  
&=& \frac{1}{2}(2-x_g)(x_g-x_H)+x_g'-3(1-\ode) \ge0\ . 
\eea 
This gravity theory has no slip since there is no 
nonminimal coupling and so the modified Poisson equations 
will multiply Newton's constant by a time varying 
factor 
\be 
\geff=\gm=\gl=1+\frac{x_g^2}{x_g(2-x_g)+2x_g'}\ . 
\ee

Finally, the field equation determines $x_\phi$ through 
\be  
-\al x_\phi = x_k\,[6n-(1-2n)\,ru]+\frac{3x_g}{2}\,\left(1-x_k-\frac{\Omega_r}{3}+\frac{x_g u}{6m}\right)-x_g u\left(\frac{1-m}{m}+\frac{u}{6m}\right)\,,  \label{eq:alxphi} 
\ee 
where $\Omega_r$ is the fractional energy density of radiation, a 
contributor to $\wtot$. 
Note that all the above expressions reduce to the appropriate 
shift symmetric equations (see e.g.\ \cite{Calderon:2026vgr}) in the 
limit $u=0$. 

Since in this model charge is essentially conserved 
at early times, the same forms of initial conditions as in 
\cite{paper1} apply. 
(Note the $XG_{3\phi}$ contribution to the dark energy density, and the 
equivalent contribution to the charge density, at early times is 
small compared to the $H\dot\phi g$ term since $X/(H^2\mpl)\ll1$.) 
In particular, the combination of the no ghost 
and stability conditions determine that neither the kinetic nor braiding 
contributions to the dark energy density can completely dominate early, 
that is $f\equiv x_{g,i}/\odei$ is neither 0 nor 1, 
but instead there must be a balance between them 
(and in the charge density as well). 

The early time effective charge conservation gives 
$C\sim a^{-3}$, and as in \cite{paper1} the balance between 
kinetic and braiding terms then implies 
\bea   
w&=&\frac{1}{2n-1}\\ 
m&=&\frac{(2n-1)}{6}\ , 
\eea 
evaluated in the radiation dominated era with $\ode\ll1$. (Note that 
\cite{paper1,Calderon:2026vgr} took initial conditions in the matter dominated era and so have $m=(2n-1)(1-w_{{\rm tot},i})/4=(2n-1)/4$ instead.)

\subsection{Dynamical System Analysis} \label{sec:dynsys} 

Much of the dynamical system analysis carried out in 
\cite{Calderon:2026vgr} can be generalized, now to a 3D phase space 
of $(x_k,x_g,u)$, i.e.\ on the $\Omega_r=0$ and $\Om=0$ (invariant) faces of the flow. 
We discuss this in detail in Appendix~\ref{sec:apxdyn} and here 
give just a few general results. 
Properties following from the \emph{form} alone of the dynamical 
system in Eq.~(\ref{eq:ode_autonomous}) include: 
\begin{itemize} 
\item As in the shift symmetric case, $x_k'\propto x_k$ and 
$x_g'\propto x_g$ so each of these variables is restricted 
to its initial sign. 
\item Further, $u'\propto u$ so also $u$ is restricted to its 
initial sign. Thus the 3D phase space is restricted to one 
octant. Symmetry breaking is a growing direction, that is the system 
evolves away from early time $u=0$. 
\end{itemize} 

Since we are particularly interested in the phantom divide 
crossing, note that 
the crossing point $w=-1$ gives the equation 
\be 
2nx_k=x_g\left(x_\phi-1+\frac{u}{3m}\right)\ , \label{eq:wcross} 
\ee 
generalizing Eq.~(6.1) of \cite{paper1}.  
To cross the phantom 
divide the ratio $x_k/x_g$ has to evolve in such a way as to 
bring $w$ from one side of $w=-1$ to the other. 
Unfortunately the parabola $Q(x_k,x_g)$ analyzed in 
\cite{Calderon:2026vgr} to determine phantom crossing analytically 
becomes much more complicated when the field dependence 
generates $u\ne0$. 
Basically the analysis indicates that one needs opposite signs for $u$ and $r$, with $r\le 0$ (so the braiding term can grow relative to the kinetic term) and $u>0$. This is borne out by numerical solutions achieving phantom crossing. 

Appendix~\ref{sec:apxdyn} presents further analysis concerning fixed points, eigenvalues, etc.

\section{Radiation Era, Matter Era, and Final States for Dark Energy} \label{sec:wfinal}

\subsection{Radiation and Matter Dominated Eras} 

During the radiation and matter dominated eras, when 
$\ode\ll1$ and shift symmetry still holds substantially, 
one can obtain analytic expressions for certain quantities 
of interest. These follow from charge conservation and balance between the charge terms, applied at the initial conditions in the radiation dominated era. We have 
\bea 
x_\phi&=&\frac{3[1+\wtot(a)]}{4(1-2n)}\\ 
w&=&\frac{-1+f(a)\,[1-3\wtot(a)]/4}{1-2n}\ .
\eea 
These hold even for the transition between radiation and 
matter domination. Recall that $f\equiv x_{g,i}/\odei$ 
is just a number, and only in this subsection have we 
generalized it to a function $f(a)=x_g/\ode$. 

The above expressions reduce to  
\bea 
\begin{aligned}
&\text{Radiation Domination:}\quad &x_\phi&=\frac{1}{1-2n}\ , &w&=\frac{-1}{1-2n}\ ,\\
&\text{Matter Domination:}\quad &x_\phi&=\frac{3}{4(1-2n)}\ , &w&=\frac{-1+n/2}{1-2n}\ , &f=2n&\ .
\end{aligned} 
\eea 
Note that regardless of what (healthy) $f$ is chosen 
initially in the radiation dominated era, $f(a)$ is attracted 
in the matter dominated era to the fixed value $2n$, so 
the initial value is moot. 
Again, Appendix~\ref{sec:apxdyn} derives this. 

\begin{figure}
    \centering
    \includegraphics[width=\linewidth]{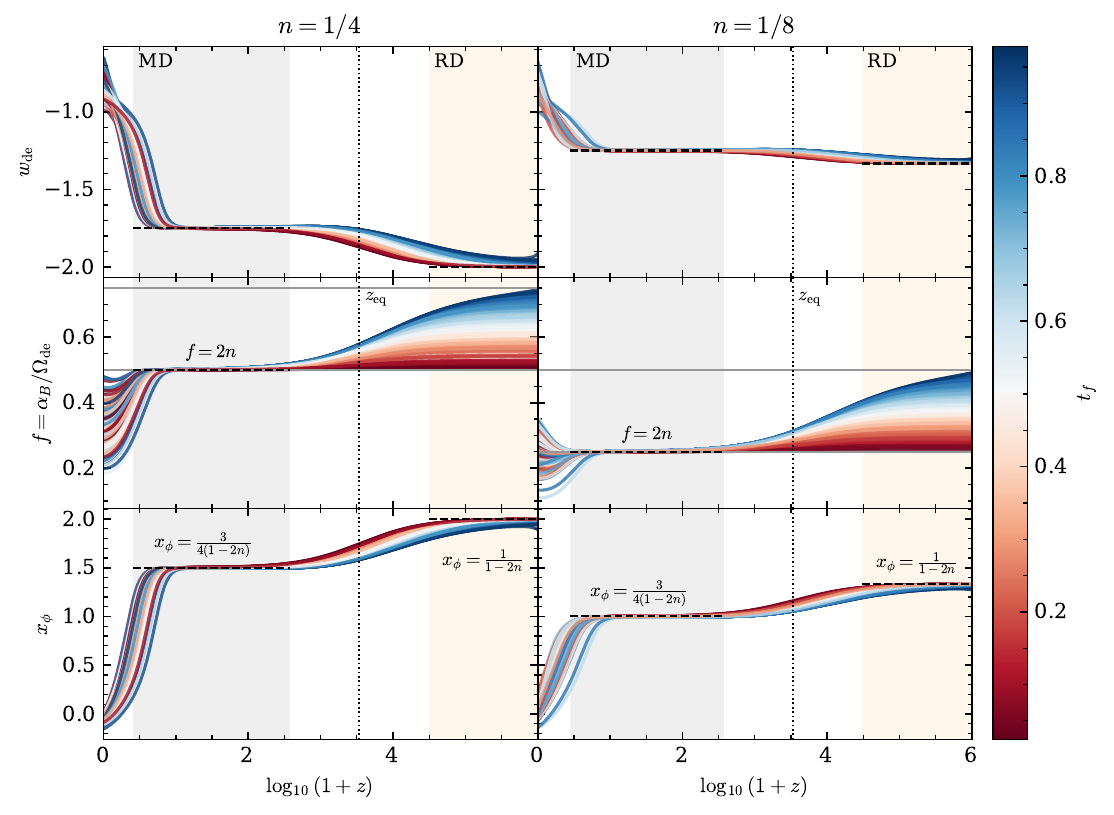}
    \caption{
    We show $2000$ viable SBSS histories drawn from the chains, color-coded by the (rescaled) braiding fraction $t_f$. All histories converge at matter domination, where the trajectories ``forget'' their initial $f$ and the late-time phenomenology is uncorrelated with the $t_f$ values. 
    Horizontal dashed lines show the radiation dominated (RD) and matter dominated (MD) era analytic values.}
    \label{fig:histories_f}
\end{figure}

Fig.~\ref{fig:histories_f} illustrates that the numerical evolution indeed follows 
these analytic behaviors, and in particular that 
the late time behavior is uncorrelated with the 
initial value of $f$. 
When we constrain the model with the data we will find 
that the posterior for $f$ is flat within the 
healthy region, and so it is not actually needed 
as a parameter.

\subsection{Final Fate of Dark Energy} \label{sec:intu} 

To investigate the asymptotic future state of the dark 
energy cosmology, we can carry 
out an analytic evaluation of the final states (future fixed 
points). Consider the dark energy 
density evolution equation coming from the continuity equation, 
\be 
\ode'=3\ode(\wtot-w)\ . 
\ee 
This gives three possible 
end states: 1) $\ode\to0$, i.e.\ dark energy vanishes 
(as with $\wl>0$), 
2) $\wtot\to \wl$ due to dark energy domination, so $\ode\to1$  
(necessarily with $\wl<0$), 
or 3) $\wtot\to \wl$ due to dark energy tracing the dominant 
background component, i.e.\ matter at late times. Then dark 
energy and matter scale together, with $\wl\to0$ and some finite 
value for $\ode$. 

If dark energy density vanishes, $\ode=0$ and $\wl>0$, then 
Eq.~(\ref{eq:odeu}) gives $x_k$ proportional to $x_g$. If 
$x_g=0$ then $x_k=0$ and the modified gravity asymptotically 
becomes simply general relativity with matter only. 
This is the matter dominated fixed point of 
\cite{Calderon:2026vgr}. 
If instead $x_g\ne0$ 
then the stability criterion $\al c_s^2=-(x_g/2)(1+x_g)\ge0$ 
is violated, taking into account that $x_H=-3$ in matter 
domination. Thus the only healthy end state with $\wl>0$ is 
$(x_k,x_g)=(0,0)$. 

If dark energy traces matter asymptotically, $\wl=0$ and finite 
$\ode$, then again $x_k$ is proportional to $x_g$, so we cannot 
have $x_g=0$. We again have $x_H=-3$ and find $x_\phi=-1/2$, 
$x_k=0$, and $\ode=x_g/2$. 
This is a new fixed point with $u=3m$, 
not appearing in the shift symmetric case. 
This asymptotic scaling cosmology will be asymptotically 
both ghost free and stable if $2(1-m)\le x_g\le 2$. 
The value of $x_g$ is determined by 
initial conditions (and constrained by, e.g., requiring $\odet\approx0.7$). However, the condition $2(1-m)<2$ 
requires $m>0$ so $n>1/2$, meaning early matter domination 
will be violated, so this is not a viable end state for our 
cosmology. 

For the dark energy domination case, $\ode=1$ and $\wl<0$, the only 
healthy case is where $x_k\ne0$ and $x_g\ne0$, and this ends up 
in a de Sitter state. 
Appendix~\ref{sec:apxdyn} presents a more sophisticated analysis. 

The three final states are summarized in Table~\ref{tab:wend}. 
Dark energy vanishing corresponds to the fixed point M; 
the new scaling solution is the generalization of fixed 
point B with constant $u=3m$, and dark energy dominating 
can only be the fixed point dS.

\begin{table}[tb]
\centering
\begin{tabular}{l|cccccccl}
\toprule 
Dark Energy & $\ode$ & $w_{\rm fut}$ & $x_k$ & $x_g$ & $u$ & $\al$ & $c_s^2$ & Notes \\ 
\midrule 
Vanishes & 0 & $>0$ & 0 & 0 & - & 0 & - & Restores to GR \\  
Scales & $x_g/2$ & 0 & 0 & $x_g$ & $3m$ & $\frac{3x_g[x_g-2(1-m)]}{2}$ & 
$\frac{2-x_g}{3[x_g-2(1-m)]}$ & $2(1-m)\le x_g\le2$ \\ 
Dominates & 1 & $-1$ & $-1$ & $2n$ & 0 & $2n(2-n)$ & $\frac{1-n}{2-n}$ & dS ; $0<n\le1$ \\ 
\end{tabular} 
\caption{Future asymptotic states of dark energy and the conditions 
to be healthy in that limit. Note that $m=0$ implies $g=0=x_g$ and so 
is excluded except for dark energy vanishing. 
} 
\label{tab:wend}
\end{table}


\section{Constraints from Current Data} \label{sec:results} 

\begin{figure}
    \centering
    \includegraphics[width=1\textwidth]{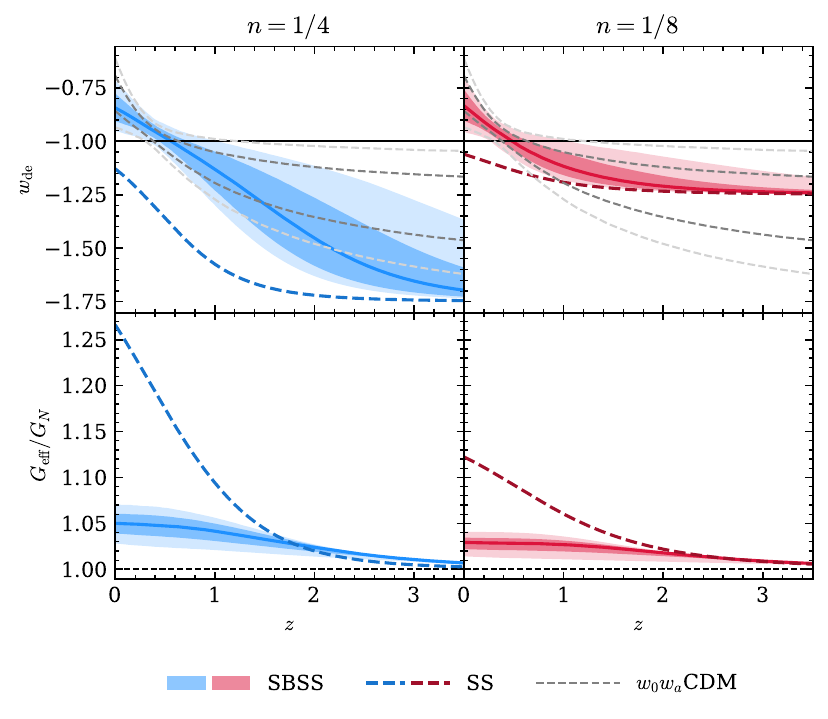}
    \caption{[Top panel] Effective dark energy equation of state $\wde(z)$ for $n=1/4$ (left) and $n=1/8$ (right), for the Softly Broken Shift Symmetric (SBSS) and Shift Symmetric (SS) cases, 
    compared to \wowa. [Bottom panel] Corresponding effective gravitational coupling strength $\geff(z)$ 
    in units of Newton's constant. 
    Solid curves are the median and shaded bands are 68.3\% and 95.4\% confidence levels. All histories converge to the shift symmetric behaviour in the past. 
    }
    \label{fig:wz12525}
\end{figure} 

We constrain the model and other cosmological 
parameters using current data corresponding to distances 
measured from baryon acoustic oscillations (BAO), from 
DESI DR2 \cite{desi2}, distances measured from Type Ia 
supernovae (SN), from the Union3.1 compilation \cite{Hoyt:2026fve}, 
and the full Planck 2018 (PR3) power spectra from the cosmic microwave 
background \cite{Planck:2018nkj,Planck:2019nip,Balkenhol:2024sbv}. 
The posterior medians, best fits, and uncertainties for the parameters, as well as the model $\chi^2$, are 
computed with \textsc{cosmonuts} \cite{Calderon:cosmonuts}, using the nested slice sampling algorithm \cite{yallup2026nestedslicesamplingvectorized} implemented in \texttt{BlackJAX} \cite{cabezas2024blackjaxcomposablebayesianinference}. 

The power law index $n$ in the action (and hence $m$ as well) 
is fixed, not sampled over, 
and we present results for two different cases: $n=1/4$ 
(giving $\we=-2$) and $n=1/8$ (giving $\we=-4/3$). 
Recall that $m$ is determined from $n$ through early 
charge conservation. 
The field dependence scale ratio $r\equiv \lambda_k/\lambda_g$ is sampled over, as is $u_{\rm ref}$, 
the logarithmic derivative of the potential (related to 
the initial value of the scalar field) at a reference 
time. Note that 
given the three dynamical evolution 
equations there are three initial conditions, but those 
for $x_k$ and $x_g$ 
can be related to the dark energy density fraction 
today, $\odei=1-\Om$, and the initial braiding fraction $f$; 
recall the initial $f$ is moot as $f(a)$ goes to an 
attractor value during matter domination. Thus there 
are two effective free parameters for the dark energy: $r$ and $u_{\rm ref}$ 
-- the same number as in \wowa. 
In Appendix~\ref{app:priors}, we discuss our choice of sampling parameters and priors, and study their impact on the Bayesian analyses providing 
log evidence (Bayes factor) to supplement $\chi^2$ by performing profile likelihood analyses of the different model cases. 

Figure~\ref{fig:wz12525} illustrates the $w(z)$ behavior, showing the median fit and  
the 68.3\% and 95.4\% uncertainty bands, and comparing to 
the \wowa\ case. The $n=1/8$ case in particular is quite similar to 
the \wowa\ cosmology for $w(z)$. Note that SBSS moderates  
the behavior of the corresponding unbroken shift symmetry 
case (which for these constant $n$ does not cross the 
$w=-1$). 
SBSS does produce an effective dark 
energy equation of state that crosses the phantom divide, 
rising from its fixed, shift symmetric, charge conserving 
value of $w[{\rm mat}]=(-1+n/2)/(1-2n)$ during matter domination, through $w=-1$ 
at $z\approx 0.5$, to values today $w_0\approx -0.8$. 
Of course since SBSS is 
a modified gravity theory it will also give a different 
prediction for growth of structure and gravitational 
lensing than general relativity (GR). The plotted results for $\geff(z)$ show modest deviations from 
GR.

\begin{figure}
    \centering
    \includegraphics[width=\textwidth]{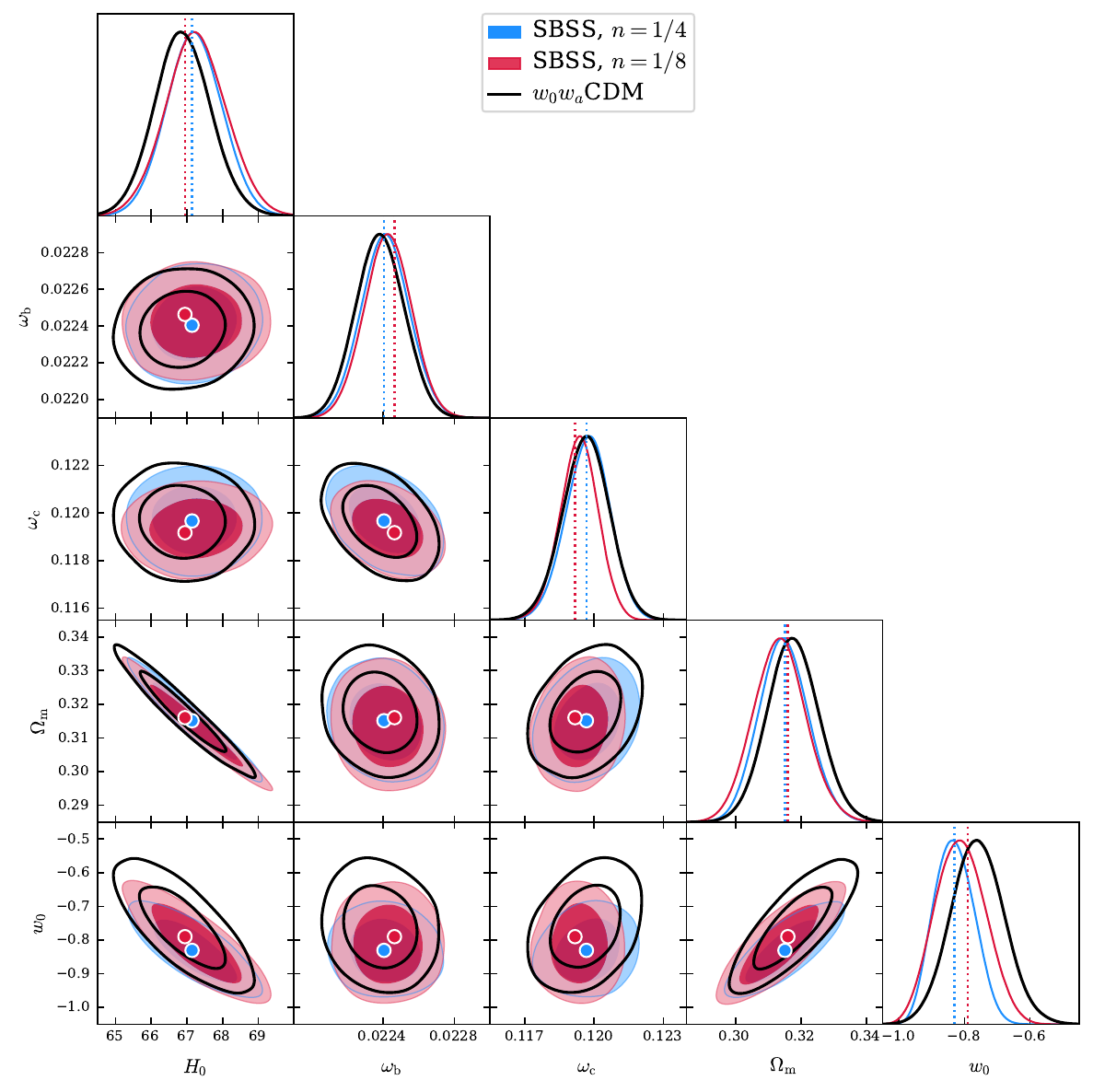}
    \caption{ Triangle plot at 68.3\% and 95.4\% posterior 
    probability for the standard cosmological parameters, for SBSS with $n=1/4$ and $n=1/8$, compared to \wowa. Best fit points for SBSS cases are indicated by circles. 
    }
    \label{fig:tri125}
\end{figure}

\begin{figure}
    \centering
    \includegraphics[width=\textwidth]{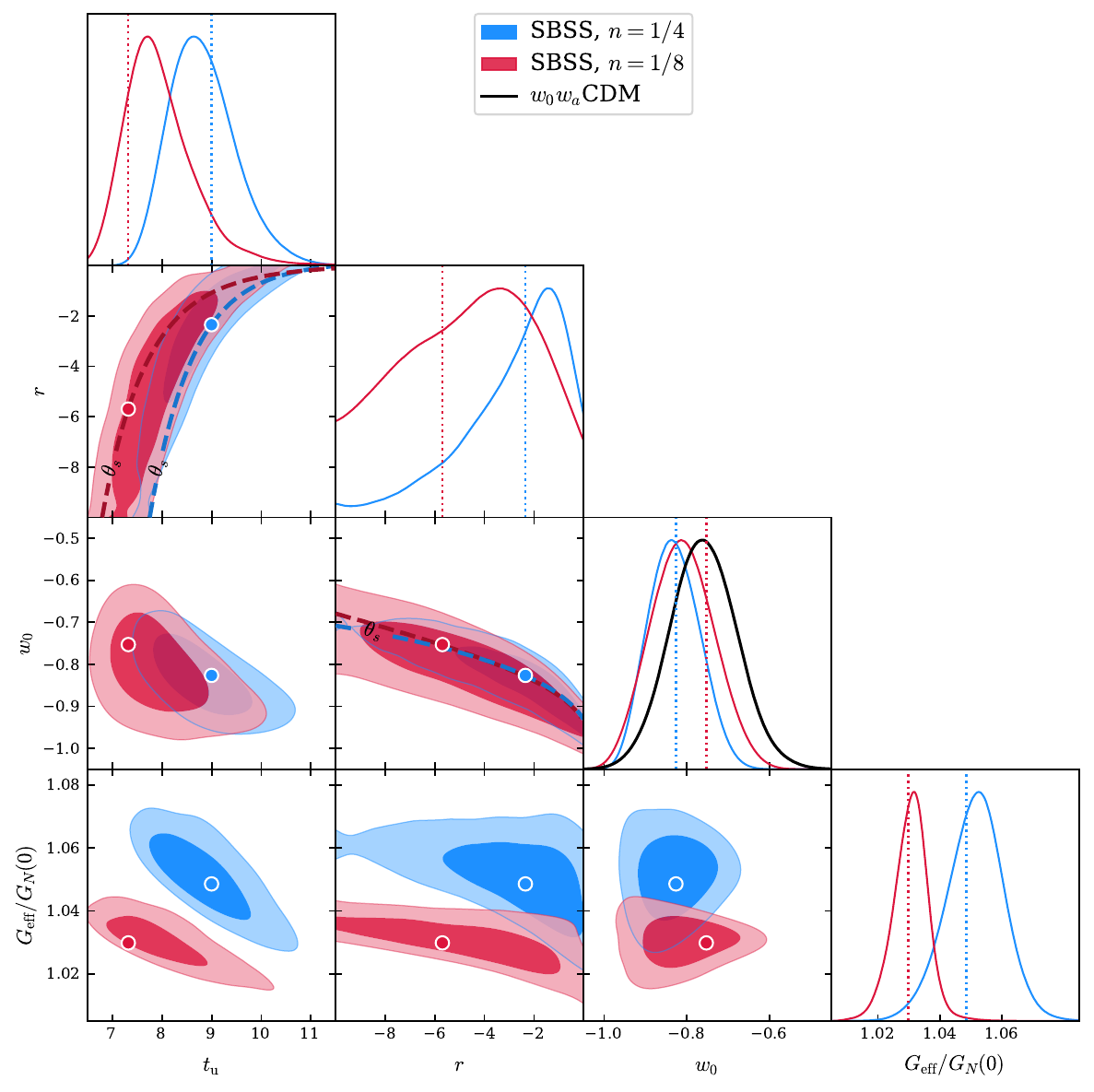}
    \caption{Triangle plot at 68.3\% and 95.4\% posterior 
    probability for the SBSS model parameters, for $n=1/4$ and $n=1/8$. Best fit points are indicated by circles. Dashed thick lines in the $r$ vs $t_u$ and $r$ vs $w_0$ panels denote isocontours of constant $\theta_s$, i.e.\ the mirage degeneracy line of $w_0$-$w_a$ \cite{Linder:2007ka}.
    }
    \label{fig:tri25}
\end{figure}

Figure~\ref{fig:tri125} and Figure~\ref{fig:tri25} show the 
standard cosmological parameter and model parameter constraints, respectively, for the 
$n=1/8$ and $n=1/4$ cases. Note the standard 
cosmological parameters are in good agreement with the 
values in \wowa, and the uncertainties are quite comparable.

\begin{figure}
    \centering 
    \includegraphics[width=0.75\linewidth]{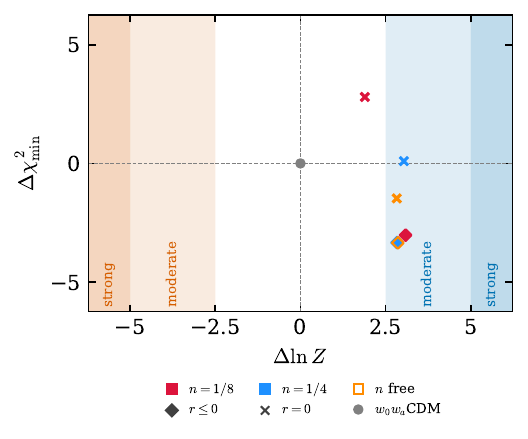}
    \caption{Improvements in fit to data in terms of $\Delta\chi^2$ and Bayes factors $\Delta\ln Z$ for the various SBSS scenarios, relative to the \wowa\ model (with $\Sigma m_\nu=0$).} 
    \label{fig:evidence}
\end{figure}

Table~\ref{tab:gof_pr3} 
presents the goodness of fit 
for various SBSS cases in terms of total $\Delta\chi^2$ 
with respect to \wowa, and the breakdown by observational 
probe. The number of effective parameters for describing 
the dark energy sector is indicated by $N_{\rm par}$, 
with standard SBSS having the same number as \wowa. 
Note that removing field dependence from the kinetic 
term by setting $r=0$ lessens $N_{\rm par}$ by one, but 
at the cost of 3--6 in $\chi^2$. However, the $r=0$ case 
can still be competitive with \wowa. 
We also show the log Bayes factor $\Delta\ln Z$, the difference 
of the log evidence for each model vs \wowa.  These 
generally show moderate preference over \wowa. 
Figure~\ref{fig:evidence} exhibits the model comparison -- one of the main results of this work. 

While we have not included growth of structure data for 
the constraints, Appendix~\ref{sec:apxgrow} shows the 
behavior for some growth quantities, which indicates that 
SBSS should do about as well as \wowa\ there as well. 

Thus, SBSS provides a physical foundation for crossing the 
phantom divide, and can do so in not only a healthy and 
viable manner but fit the current data better than \wowa. 
The $\chi^2$ improvement is $-3$, and the log Bayes factor 
improvement is $3$, generally regarded as moderate 
preference.


\begin{table*}
\centering
\small
\begin{tabular}{lcccc>{\columncolor{tabcol0}}cc}
\toprule
 &  & \multicolumn{4}{c}{$\Delta\chi^2$} &  \\
\cmidrule(lr){3-6}
Model & $N_{\rm par}$ & DESI DR2 & Union3.1 & CMB & \bf{Total} & $\Delta\ln Z$ \\
\midrule
$w_0w_a$CDM & $2$ & $8.62$ & $26.66$ & $999.67$ & $1034.95$ & -- \\
\midrule
SBSS, $n=1/8$ & $2$ & $-0.19$ & $+0.10$ & $-2.93$ & $-3.02$ & $+3.09\pm0.09$ \\
SBSS, $n=1/4$ & $2$ & $-0.26$ & $+0.51$ & $-3.58$ & $-3.33$ & $+2.84\pm0.09$ \\
\midrule
SBSS, $n=1/8$, $r=0$ & $1$ & $+1.85$ & $+2.94$ & $-1.99$ & $+2.80$ & $+1.89\pm0.10$ \\
SBSS, $n=1/4$, $r=0$ & $1$ & $+0.55$ & $+1.75$ & $-2.21$ & $+0.10$ & $+3.04\pm0.07$ \\
\midrule
SBSS, $n$ free, $r=0$ & $2$ & $+0.52$ & $+1.55$ & $-3.54$ & $-1.47$ & $+2.83\pm0.13$ \\
SBSS, $n$ free, $r$ free & $3$ & $-0.33$ & $+0.49$ & $-3.51$ & $-3.35$ & $+2.86\pm0.11$ \\
\bottomrule
\end{tabular}
\caption{
Goodness of fit and Bayesian evidence with DESI DR2 BAO, Union3.1, and the full Planck 2018 (PR3) CMB likelihoods (low-$\ell$ TT, low-$\ell$ EE and plik-lite TTTEEE; the CMB column is their sum), relative to $w_0w_a$CDM. The $w_0w_a$CDM row shows its best-fit $\chi^2$; every other row $\Delta\chi^2=\chi^2-\chi^2_{w_0w_a}$ at its own best fit and $\Delta\ln Z=\ln Z-\ln Z_{w_0w_a}$. $N_{\rm par}$ counts the parameters beyond $\Lambda$CDM. Our baseline SBSS 
has $t_u$ and $r\leq0$ free.  
The evidence $\ln Z$ is the mean of 3 nested-sampling runs with different seeds, with error $\sigma_{\rm seeds}/\sqrt{3}$, added in quadrature to the errors of the \wowa\ runs (also from 3 different seeds). 
Note that all of these assume massless neutrinos and priors reported in \Cref{tab:priors}.
}
\label{tab:gof_pr3}
\end{table*}


\section{Conclusions} \label{sec:concl} 

Cosmic acceleration is a fundamental mystery in cosmology, 
with a requirement for new physics beyond the standard model 
of particle physics. On top of this, current data point toward 
a variation of the acceleration/deceleration corresponding 
to an effective dark energy evolving across $w=-1$, the 
phantom divide. Rather than viewing this as an added puzzle, 
one can consider it an important clue to the physics required. 

We consider a model maximizing the principles of simplicity, 
robustness or protection from high energy quantum effects, 
and naturality: we avoid any nonminimal couplings and any 
cosmological constant or potential, and employ Horndeski 
gravity shift symmetry 
to the extent possible -- only softly breaking it, and that 
with exponential field dependence terms. This softly broken 
shift symmetry (SBSS) modified gravity is remarkably 
successful. 

Fitting data from baryon acoustic oscillation and supernova 
distances, and CMB power spectra, yields an effective dark 
energy equation of state very similar to the phenomenological 
\wowa, but with a physical action as a foundation. Indeed, 
the goodness of fit is better than \wowa\ 
(by $\Delta\chi^2\approx -3$), with the same number of 
effective free parameters. In terms of log Bayes factor, 
$\Delta\ln Z\approx3$ with respect to \wowa. Moreover, 
the gravitational strength deviations from general relativity 
are modest, but testable. 

SBSS is viable (describing a universe like the cosmology 
we observe) and healthy (free of 
ghosts and instability) in the relevant regime. Arising from a well defined action, 
one can calculate any cosmological observable as well as 
one can in general relativity. Appendix~\ref{sec:apxgrow} 
shows some results 
investigating fits to cosmic growth of structure quantities. 

Such a simple theory, with a well-founded physical 
foundation and highly predictive, together with attention 
paid to early universe (high energy) quantum protection, 
is a promising avenue for exploring new physics for 
cosmic acceleration. The success in fitting observational 
data shows SBSS is not only an interesting theory, but one 
capable of describing {\it our\/} universe.

\section*{Acknowledgments}

We thank the Galileo Galilei Institute for hospitality during part of this work, and lively discussions, especially with Jeremy Sakstein. 
We also thank Tessa Baker and James Hallam for useful discussions. 
R.C.\ is funded by the Czech Ministry of Education, Youth and Sports (MEYS) and European Structural and Investment Funds (ESIF) under project number CZ.02.01.01/00/22\_008/0004632. 
E.L.\ thanks the Simons Foundation for support as a Simons Visiting Scientist at GGI. We acknowledge the use of the Boltzmann solver \texttt{hi\_class} \cite{Zumalacarregui:2016pph}, which was modified to include our model and compute the CMB spectra.

\bibliographystyle{JHEP}
\bibliography{refs}

\appendix

\section{Field Redefinitions} \label{sec:redef}

The action given in Eq.~(\ref{eq:actionkgsep}) has two functions 
of the field $\phi$ and two functions of its derivative, i.e.\ $X$. 
Can we remove one function of $\phi$ through a field 
redefinition, either from the kinetic term (possibly even 
making it of canonical form) or the braiding term? We show 
here that the answer is basically no: any such field redefinition 
will in turn generate a new kinetic term with its own $\phi$ 
dependence -- we do not reduce the number of functions. 

To sum up the results obtained below, trying to remove field 
dependence in one action term yields: 
\bea 
V_k(\psi)\,\kappa(\dot\psi)-V_g(\psi)\,\gamma(\dot\psi)\,\Box\psi 
\quad&\to&\quad \tilde V_k(\phi)\,\tilde\kappa(\dot\phi)+V_{\rm new}(\phi)\,\dot\phi^2\tilde\gamma(\dot\phi)-\tilde\gamma(\dot\phi)\,\Box\phi  \ ,\label{eq:g3can}\\ 
V_k(\psi)\,\kappa(\dot\psi)-V_g(\psi)\,\gamma(\dot\psi)\,\Box\psi 
\quad&\to&\quad \tilde\kappa(\dot\phi)+V_{\rm new}(\phi)\,\dot\phi^2\tilde\gamma(\dot\phi)-[\tilde V_g(\phi)/\phi]\,\tilde\gamma(\dot\phi)\,\Box\phi\ . \label{eq:kcan} 
\eea 
(The specific forms of the new field dependence depends on the 
original functions; Eq.~\ref{eq:kcan} is for monomial $\kappa$ and 
exponential $V_k$.) 
That is, removing the field dependence from one term causes a new 
term to appear with its own field dependence: no overall reduction 
occurs. 

To demonstrate this explicitly, take a field redefinition $\phi=\phi(\psi)$. Then 
\be 
\Box\phi=(\phi_\psi \psi_j)_j=\phi_\psi\,\Box\psi-2Y\phi_{\psi\psi}\ , \label{eq:redefbox} 
\ee 
where the new kinetic variable is 
$Y\equiv (1/2)\dot\psi^2$. 

The braiding term becomes 
\be 
-G_3\,\Box\phi = -\gamma(X(Y))\,V_g(\phi(\psi))\,\phi_\psi(\psi)\,\Box\psi+2V_g(\phi(\psi))\,\phi_{\psi\psi}(\psi)\,Y\gamma(X(Y))\ ,  
\ee 
While we can normalize the box term to leave only the dependence 
on the new field derivative term $Y$, without a prefactor 
dependent on $\phi$, this still gives a new 
field dependent kinetic term, 
\be 
K^{\rm new}=2V_g(\phi(\psi))\,\phi_{\psi\psi}(\psi)\,Y\gamma(X(Y))\ . \label{eq:knew} 
\ee 

Let's give a specific example where $\gamma(X)=X^m$. Then 
$\gamma(X(Y))=(\phi_\psi)^{2m} \,Y^m$, and the braiding normalization 
condition is 
\be 
(\phi_\psi)^{2m+1}\,V_g=1\ , 
\ee 
i.e.\ the field redefinition is given by $\psi=\int d\phi\,V_g(\phi)^{1/(2m+1)}$. The new kinetic contribution is 
then 
\be 
K^{\rm new}=2\,\frac{\phi_{\psi\psi}}{\phi_\psi}\,Y^{m+1}\ . 
\ee 
The full action after field redefinition becomes 
\be 
{\mathcal L}=\frac{1}{2}\,R+V_k(\phi(\psi))\,\kappa(X(Y))
+2\,\frac{\phi_{\psi\psi}}{\phi_\psi}\,Y^{m+1} 
-Y^m\,\Box\phi\ . \label{eq:renomg}
\ee 
We are still left with two functions of the field derivative 
-- $Y^m$ coming from $\gamma$, and $\kappa$ -- and two functions 
of the field value -- $V_k$ and $\phi_{\psi\psi}/\phi_\psi$, 
essentially coming from $V_g$. 

We do note that the particular choice of $m=-1$ plus the 
braiding normalization has the 
consequence of producing a purely field dependent term, 
i.e.\ a potential $V^{\rm new}=-2\phi_{\psi\psi}/\phi_\psi$, 
in the new kinetic term. Conversely, 
for the particular field redefinition $\phi\sim e^{\lam\psi}$, 
i.e.\ $V_g\sim e^{-(2m+1)\lam\phi}$, there is no field dependence 
in the new kinetic term. If one adopts both the exponential 
field redefinition and $m=-1$ then the generated new contribution 
to the kinetic term is a cosmological constant. 

An equivalent generation of a new kinetic term results from 
choosing a field redefinition to remove the field dependence 
of the kinetic term. If $\kappa(X(Y))=X^n$ then we can remove 
the $V_k(\phi)$ dependence by choosing $(\phi_\psi)^{2n}\,V_k=1$. 
But Eq.~(\ref{eq:redefbox}) still applies and we generate 
a new kinetic term as in Eq.~(\ref{eq:knew}) that depends 
on both field value and derivative.

\section{Stability of the ODE system} \label{sec:apxdyn}

Here, we study the fixed points of Eq.~\ref{eq:ode_autonomous} and their stability. Note that the matter fraction is $\Omega_m=1-\ode-\Omega_r$. The
hyperplanes $x_k=0$, $x_g=0$, $u=0$, $\Omega_r=0$ (the matter face) and $\Omega_m=0$ (the radiation
face) are invariant. The shift-symmetric flow ($u=0$) studied in \cite{Calderon:2026vgr}\textemdash and generalized here to include radiation\textemdash is therefore three-dimensional and bounded,
$0\le\Omega_r\le1-\ode$. At early times $\ode\to0$, and the background equation of state is $w_{\rm tot}=1/3$. To leading
order in $\ode$, we get
\begin{equation}
f'=-\frac32\,(1-\Omega_r)\,\frac{f(1-f)(f-2n)}{6n-(4n+1)f}\ .
\label{eq:rb_fprime}
\end{equation}
Deep in radiation ($\Omega_r=1$) every ray of constant $f$ is an
orbit, so the radiation point $R=(x_k,x_g,\Omega_r)=(0,0,1)$ is a star (repeller): histories leave it along a
ray labelled by an arbitrary constant $f_i$, with
\begin{equation}
\gamma\equiv\frac{d\ln\ode}{d\ln a}=\frac{2(2-n)}{1-2n}\qquad \Rightarrow
\qquad
\wde=\frac{1}{2n-1}\ ,
\label{eq:rb_R}
\end{equation}
independently of the initial $f_i$. Once matter dominates, Eq.~(\ref{eq:rb_fprime}) drives every history to the 
ray $f=2n$ as $f-2n\propto a^{-3/4}$.  
There $\gamma=3(2-n)/[2(1-2n)]$ and
$\wde=-(2-n)/[2(1-2n)]$. The segment $x_k=x_g=0$ from $R$ to the matter point $M$ is the background
radiation-to-matter orbit.

\begin{table}[h]
\centering
\renewcommand{\arraystretch}{1.35}
\begin{tabular}{c|c|c|c|c|c}
\textbf{Point} $(x_k,x_g)$ & $\wde$ & $(x_k,x_g)$ SS block & $\lambda_r$ & $\lambda_u$ & $n=\tfrac14$ \\ \hline
$K$\ $\big(\tfrac{1}{2n-1},0\big)$ & $\dfrac{1}{2n-1}$ & $\dfrac{n-2}{2n-1},\ \dfrac{3}{2n-1}$ &
$\dfrac{2(2-n)}{2n-1}$ & $\dfrac{3(n-1)}{2n-1}$ & $(\tfrac72,-6),\,-7,\,\tfrac92$ \\
$B$\ $(0,1)$ & $\dfrac{3}{4n+1}$ & $\dfrac{6(2-n)}{4n+1},\ \dfrac{9}{4n+1}$ & $\dfrac{4(2-n)}{4n+1}$ &
$\dfrac{3(2n-1)}{4n+1}$ & $(\tfrac{21}{4},\tfrac92),\,\tfrac72,\,-\tfrac34$ \\
dS\ $(-1,2n)$ & $-1$ & $-3,\ -3$ & $-4$ & $0$ & $(-3,-3),\,-4,\,0$
\end{tabular}
\caption{Fixed points of the linearized system~(\ref{eq:ode_autonomous}) and their associated eigenvalues $\lambda_i$, for $m=(2n-1)/6$. The SS block refers to the eigenvalues of the reduced, shift symmetric case $(x_k,x_g)$;
all other columns are for SBSS with $\lambda_r$ and $\lambda_u$  the eigenvalues in the $\Omega_r$ and $u$ directions.}
\label{tab:rb_fp}
\end{table}

Table~\ref{tab:rb_fp} summarizes the fixed points. 
de Sitter, with $f=2n$, is the unique attractor of the shift-symmetric flow. The kinetic point $K$
is a saddle. The braiding point $B$ repels within $u=0$ but attracts along $u$. A typical history
therefore runs $R\to M\to{\rm dS}$: it leaves $R$ (radiation domination) along its $f_i$ ray, bends onto $f=2n$ near $M$ (matter domination),
and ends at dS. The broken case, $u\neq0$, is slightly more involved.
The dS fixed point has a zero eigenvalue along $u$, so linear theory cannot say anything about the stability along the symmetry breaking direction $u$. Using centre manifold analysis (e.g.\ \cite{carr1981applications,bahamonde,Calderon:2019vog}), we have 
\begin{equation}
u'=c_2u^2\ ,\qquad c_2=\frac{(1-2n)(r-r^*)}{2(2-n)}\ ,\qquad r^*\equiv\frac{n-1}{m}=\frac{6(n-1)}{2n-1}\ ,
\end{equation}
so dS attracts $u$ from one side only, and the side flips at $r=r^*$ ($r^*=9$ at $n=1/4$). The system has one further fixed point, $K_u=\big(x_k,x_g,u\big)=\big(\tfrac{1}{2n-1},\,0,\,
\tfrac{6n(n-1)}{r(2n-1)}\big)$. Its $x_g$ eigenvalue is $n(r^*-r)/r$, so $K_u$ and dS exchange
stability at $r=r^*$. 
Finally, at $R$, $\lambda_u=(4n-5)/(2n-1)$ for every initial $f$, i.e.\ $u\propto a^{(4n-5)/(2n-1)}$, so the symmetry breaking is vanishingly small until late times, what we call softly broken.

\section{Analysis Details: Parameter Priors \& Profile Likelihoods}
\label{app:priors} 

The Lagrangian constants $c_k$, $c_g$, $\lambda_k$ and $\lambda_g$, together with the initial field
values, are not physical parameters on their own. A shift $\phi\to\phi+\phi_0$ is absorbed into $c_k$ and
$c_g$, and a rescaling $\phi\to s\phi$ rescales the couplings and maps $\lambda\to\lambda/s$. Their raw
values therefore depend on the normalisation of the field, and a flat prior on them would be a prior on
that convention.  We instead sample three
combinations that are invariant under these redefinitions and label a history directly; the couplings
follow from them and from the state at $a_{\rm ini}=10^{-9}$, 
in the radiation dominated era, 
with the amplitude solved for so that the
model reproduces $H^2(a=1)=H_0^2$, equivalent to the closure relation for the Friedmann expansion equation.

\paragraph{Labels and priors.}
The shared cosmological parameters are varied within their standard ranges, given in \cref{tab:priors}. Except for Planck's nuisance parameter $A_{\rm Planck}$, all priors are uniform in the sampled variables. Neutrinos are treated as massless. 
The exponent $n$ is
either fixed ($1/8$, $1/4$) or sampled in $[0.05,0.4]$. 

\begin{figure}
    \centering
    \includegraphics[width=\textwidth]{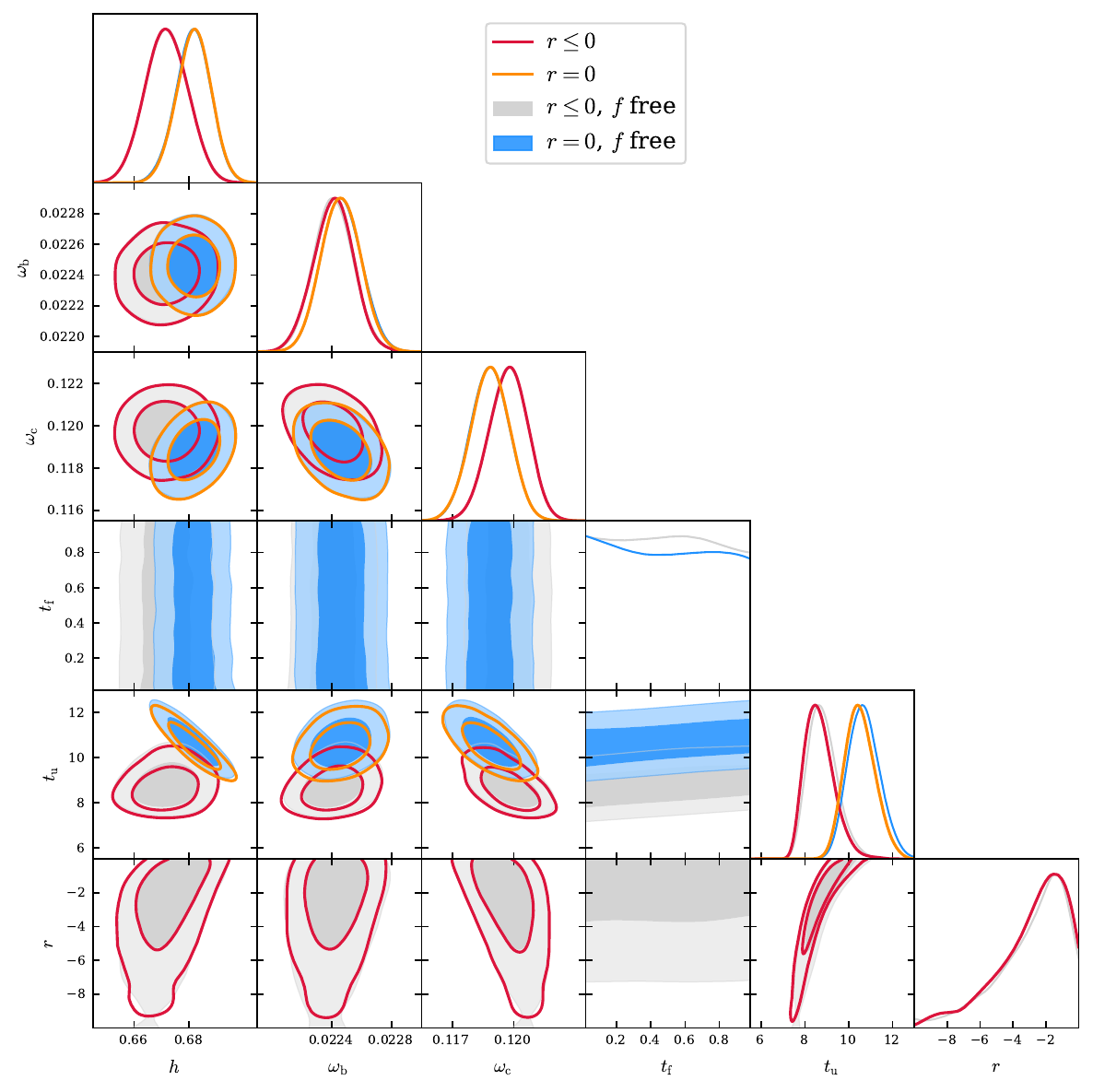}
    \caption{
    Triangle plot at 68.3\% and 95.4\% posterior 
    probability for cosmological and model parameters, for SBSS $n=1/4$, comparing our baseline fixed $f$ to allowing $f$ to range over the full healthy interval. Taking $f$ free has essentially no effect on the posteriors.}
    \label{fig:tri_full_125}
\end{figure}

\begin{itemize}
\item $f=x_g/\Omega_{\rm de}$ is frozen during radiation domination. It is stable only in the window $2n<f<6n/(4n+1)$: below it there is a gradient instability, and above it a ghost. 
For the runs where $f$ is free to vary, we take a uniform prior over the healthy window, but since the window boundaries depend on $n$, it is more convenient to define a fractional position variable that scans over the window from the lower to upper boundary. That is, 
we sample $t_f\in[0.02,0.98]$ where 
  \begin{equation}
    f=2n+t_f\left(\frac{6n}{4n+1}-2n\right).
  \end{equation}
Every draw in $t_f$ is then initially stable, for any $n$. 
For reasons discussed in the main text, our baseline fixes $f=6n(1-n)/(2+n-4n^2)$, corresponding to 
$c_s^2(a_{\rm ini})=1$. Figure~\ref{fig:tri_full_125} demonstrates 
that indeed posteriors are basically  unaffected by any 
healthy value of $f$; in particular the 1D posterior for 
$f$ is flat. 
\item $u=-\lambda_g\,{\rm d}\phi/{\rm d}N$ is the slope the field actually feels. It is independent of the
  normalisation of $\phi$, and it grows as $u\propto\Omega_{\rm de}^{\,s}$ with $s=(5-4n)/[2(2-n)]$. We
  quote it where $\Omega_{\rm de}=10^{-3}$, as $u_{\rm ref}$, 
  i.e.\ where dark energy first begins to have dynamical influence. Viable values span many decades and both signs,
  so we sample
  \begin{equation}
    u_{\rm ref}=10^{-7}\sinh t_u,\qquad t_u\in[0,13].
  \end{equation}

This prior is flat for $|u_{\rm ref}|\lesssim10^{-7}$, where the field has no effect on the expansion, and
log-uniform above that. 
To give a sense of amplitudes, note that $t_u=9$ gives $u_{\rm ref}=4\times 10^{-4}$ and $t_u=13$ gives $u_{\rm ref}=0.022$. 
Unbroken shift symmetry models have $u_{\rm ref}=0$.
\item $r=\lambda_k/\lambda_g$ is already invariant, so it is sampled directly: $r\in[-10,0]$ for the baseline results, or fixed to
  $r=0$ where indicated, or $r\in[-10,10]$ when we explore free $n$.
\end{itemize}
Points with no viable history (no $H_0$ solution, or a ghost or $c_s^2<0$ anywhere between
$a_{\rm ini}$ and today) are assigned zero likelihood.

\begin{table}
\centering
\small
\begin{tabular}{llc}
\toprule
 & Parameter & Prior \\
\midrule
Common
 & $100\theta_s$ & $\mathcal{U}[1.0355,\,1.048]$ \\
 & $\omega_{\rm b}$ & $\mathcal{U}[0.0213,\,0.0233]$ \\
 & $\omega_{\rm c}$ & $\mathcal{U}[0.112,\,0.127]$ \\
 & $\ln(10^{10}A_s)$ & $\mathcal{U}[2.904,\,3.184]$ \\
 & $n_s$ & $\mathcal{U}[0.9229,\,1.0069]$ \\
 & $\tau_{\rm reio}$ & $\mathcal{U}[0.01,\,0.125]$ \\
 & $A_{\rm Planck}$ & $\mathcal{N}(1,\,0.0025^2)$ \\
\midrule
$w_0w_a$CDM
 & $w_0$ & $\mathcal{U}[-2,\,-0.3]$ \\
 & $w_a$ & $\mathcal{U}[-3,\,2]$ \\
\midrule
SBSS
 & $n$ & $\mathcal{U}[0.05,\,0.4]$ (or fixed) \\
 & $f_i$ & fixed: $\frac{6n(1-n)}{(2+n-4n^2}$ \quad ($c_s^2=1$ at $a_{\rm ini}$) \\
 & $t_u$ & $\mathcal{U}[0,\,13]$ \\
 & $r$ & $\mathcal{U}[-10,\,0]$ \\
\bottomrule
\end{tabular}
\caption{Priors of the sampled parameters. $\mathcal{U}[a,b]$ is uniform on $[a,b]$ and $\mathcal{N}(\mu,\sigma^2)$
Gaussian. The common parameters have the same prior in every model, so they cancel in the Bayes factors; $h$ is derived
from $\theta_s$. 
}
\label{tab:priors}
\end{table}
 
\begin{figure}
    \centering
    \includegraphics[width=\linewidth]{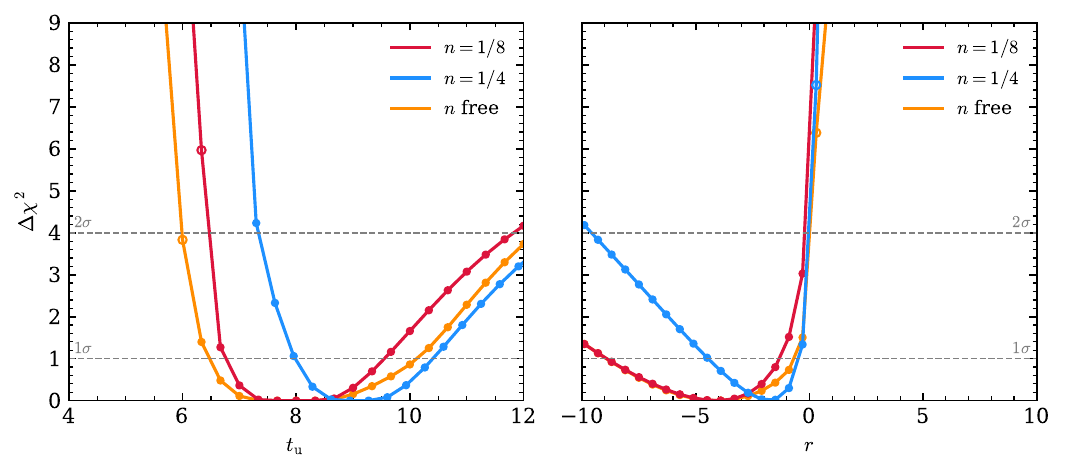}
    \caption{
    Profile likelihoods of $t_{\rm u}$ and $r$ for SBSS with $n=1/8$, $n=1/4$, and $n$ free ($n\in[0.05,0.4]$), and allowing $r\in[-10,10]$. The horizontal lines denote $\Delta\chi^2\le1$ ($1\sigma$) and $\Delta\chi^2\le4$ ($2\sigma$) limits, where every other parameter is minimised over at each fixed value. For $r>0$, $\Delta\chi^2$ roughly levels out at large (but finite) values $\Delta\chi^2>10$ not shown on the plot. This justifies our baseline analysis, restricted to $r\leq0$.  Note that these profile likelihoods, here only, use the Planck PR4 distance priors \cite{Lemos:2023xhs}. We have checked that these compressed measurements yield nearly identical posterior distributions as the full CMB spectra for the SBSS model, especially when combined with DESI DR2 and Union3.1.}
    \label{fig:profiles}
\end{figure}

\paragraph{Profile likelihoods.} To gauge the impact of our choice of priors on the Bayesian constraints presented in the main text, we perform a profile likelihood analysis for the symmetry breaking parameter $t_u$ and $r\equiv \lambda_k/\lambda_g$. The results are presented in \Cref{fig:profiles}.

\section{Impact on Cosmic Growth of Structure} \label{sec:apxgrow} 

We have compared the modified gravity model to cosmic distances 
and the CMB, and presented the deviations from general relativity 
in terms of $\geff(z)$, in the main text. Here we briefly look at 
the impact on cosmic growth observations. 

\begin{figure}
    \centering
    \includegraphics[width=\textwidth]{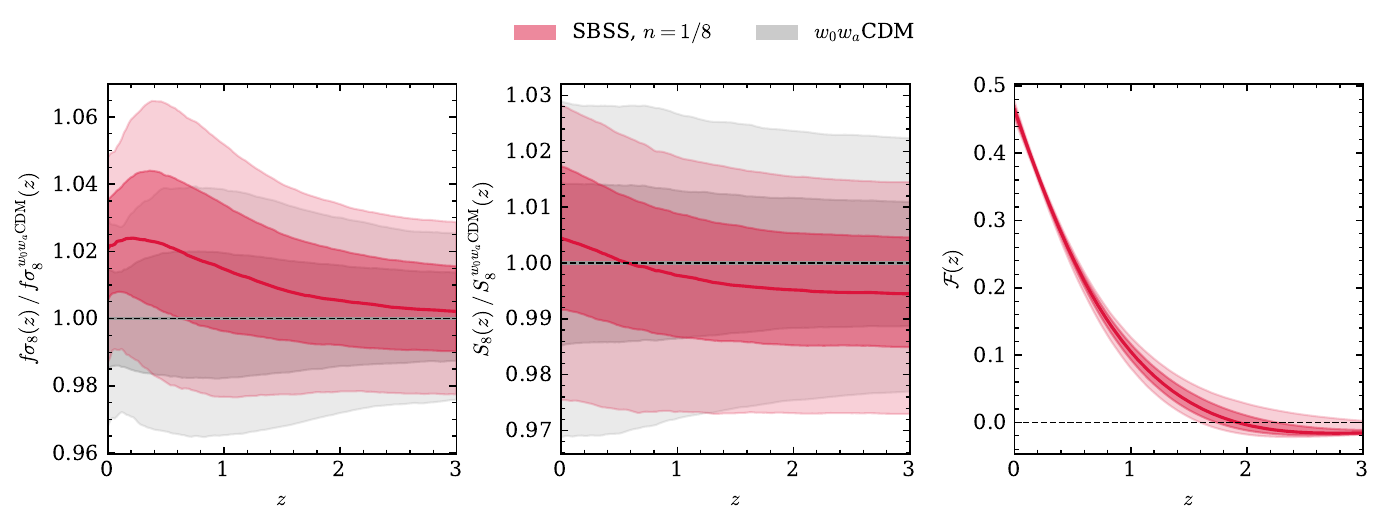} 
    \caption{
Posterior bands of cosmic growth quantities for SBSS $n=1/8$ compared to \wowa, fitted 
to DESI DR2 BAO, Union3.1 
supernovae, and the full Planck PR3 CMB likelihood. 
The left panel plots $f\sigma_8(z)$, the middle panel $S_8(z)$, and the 
right panel 
the ISW--galaxy cross-correlation kernel 
  $\mathcal{F}\equiv 1-f-{\rm d}\ln\Sigma/{\rm d}\ln a$.  
  Here $f$ is the linear growth rate and
  $\Sigma$ the quasi-static lensing gravitational coupling ($\Sigma=\geff$ in our case). Shaded regions are the 68.3\% and 95.4\% credible bands, solid lines the medians, from 2000 posterior samples.
  }
\label{fig:growth}
\end{figure}

Figure~\ref{fig:growth} shows three functions describing effects 
on growth: the logarithmic growth rate $f\sigma_8$, measurable from 
redshift space distortions, the mass fluctuation amplitude quantity 
$S_8(z)$, and the ISW-galaxy crosscorrelation kernel $\mathcal{F}(z)$. 
(Note the integrated Sachs-Wolfe contribution to the CMB power 
spectra is already included in the main article constraints.) 

For $f\sigma_8$, we see a modest increase over \wowa, expected since 
gravity is strengthened, i.e.\ $\geff>1$. For $S_8$ there is a  
slight increase at some redshifts, and slight decrease at others, 
since $S_8$ intertwines both $\sigma_8(z)$ and $\om$. 
Many cubic Horndeski gravity models run into issues with a negative 
ISW-galaxy crosscorrelation, disfavored by data (e.g.\ 
\cite{Seraille:2024beb}). SBSS has $\mathcal{F}(z)$ 
predominantly positive (in particular at $z\lesssim1.5$), with 
a higher redshift tail that can go slightly negative. We 
emphasize, however, that $\mathcal{F}(z)$ is merely the kernel, 
and it is the weighted {\it integral\/} of the kernel from $z=0$ to 
$z_{\rm survey}$ (see e.g.\ \cite{Hallam:2026qsk}) 
that should be positive to agree with data. 
SBSS appears to have no difficulty fulfilling this constraint.


\end{document}